\documentclass[12pt]{article}
\usepackage{a4wide}
\usepackage{amssymb}
\usepackage{graphicx}
\begin{document}
{\renewcommand{\thefootnote}{\fnsymbol{footnote}}
\begin{center}
{\LARGE  Non-covariance of the dressed-metric approach\\ in loop quantum cosmology }\\
\vspace{1.5em}
Martin Bojowald\footnote{e-mail address: {\tt bojowald@gravity.psu.edu}}
\\
\vspace{0.5em}
Institute for Gravitation and the Cosmos,\\
The Pennsylvania State
University,\\
104 Davey Lab, University Park, PA 16802, USA\\
\vspace{1.5em}
\end{center}
}

\setcounter{footnote}{0}

\begin{abstract}
  The dressed-metric approach is shown to violate general covariance by
  demonstrating that it cannot have an off-shell completion in which the
  correct infinitesimal relations of space-time hypersurface deformations are
  realized. The main underlying reason --- a separation of background degrees
  of freedom and modes of inhomogeneity that is incompatible with covariance
  --- is shared with other approaches such as hybrid loop quantum cosmology.
\end{abstract}

\section{Introduction}

The dressed-metric approach \cite{AAN} is an attempt to extend modified
Friedmann equations of loop quantum cosmology to perturbative inhomogeneity in
order to describe structure formation. If any such proposal is to be
consistent, it must respect general covariance in some form to guarantee that
the equations are meaningful: If general covariance is violated, the theory is
either plagued by spurious, unphysical degrees of freedom if one decides to
impose a restriced number of covariance transformations (or none at all), or
it is over-constrained if broken covariance transformations are imposed which
then identify physical solutions that are supposed to be distinct. A
non-covariant modification of a covariant theory has either too many or too
few propagating degrees of freedom, depending on how it is applied.

Since loop quantum cosmology \cite{LivRev} modifies the background dynamics of
a homogeneous universe, perturbative inhomogeneity is not guaranteed to obey
covariance conditions. However, the dressed-metric approach assumes that
classical observables and Hamiltonians can be used for inhomogeneity {\em
  without} modifications even while the background dynamics is modified such
that it may allow a bounce, a crucial ingredient in some of the developed
scenarios. In this paper, we provide the first analysis of covariance in the
dressed-metric approach, pointing out several previously overlooked subtleties
and ultimately reaching the conclusion that covariance is violated.

Several details of the technical implementation of the dressed-metric approach
obscure the issue of covariance, which is perhaps the reason why this
important issue has not been addressed yet. The approach postulates separate
quantizations for an isotropic background space-time and inhomogeneous
perturbations on it, even though the degrees of freedom of both ingredients
are interrelated in any covariant setting that obtains background and
perturbations from an expansion of a covariant theory. For instance, the
limited covariance transformations that remain in a spatially homogeneous
reduction of a covariant theory do not restrict the possible dynamics, which
can be modified at will. Homogeneous background dynamics that can be obtained
from some higher-curvature action, by contrast, is not arbitrary but subject
to conditions that implicitly ensure its descendance from a covariant theory
of this type. Once a covariant theory has been restricted to homogeneity,
however, the dynamics can be modified consistently in the homogeneous setting,
without any restrictions that would result from covariance or integrability
conditions in an inhomogeneous theory. By separating the degrees of freedom
into background and perturbations {\em before} implementing quantum
modifications, and then leaving the perturbative degrees of freedom
unmodified, the dressed-metric approach construes a setting in which the usual
covariance conditions are relaxed. This observation does not directly imply
that the approach violates covariance, but it shows that any analysis of
covariance in this approach is subtle and must be performed in detail.

While covariance itself has not yet been analyzed in the dressed-metric
approach, some transformations related to this condition have been discussed
in the seminal papers. However, these transformations, like the implementation
of degrees of freedom, act separately on background and perturbations and do
not respect the interrelated nature of these degrees of freedom with respect
to covariance. In particular, the dressed-metric approach replaces linear
perturbations of metric and extrinsic curvature, or of other fields used in
canonical gravity, by Bardeen potentials or curvature perturbations
\cite{Bardeen,CosmoPert}. Since these variables are invariant with respect to
small inhomogeneous coordinate transformations, they respect some partial form
of covariance. The homogeneous background dynamics, meanwhile, is made
invariant with respect to homogeneous time reparameterizations by using the
method of deparameterization \cite{GenHamDyn1,Blyth}, formulating homogeneous
evolution not with respect to a time coordinate but rather with respect to one
of the dynamical fields of the theory, given by a free massless scalar. The
resulting framework is formally consistent because no time coordinate is used
explicitly, and spatial coordinates can be adapted to the homogeneous
background.  In this sense, the dressed-metric approach constructs a
consistent quantum-field theory on a modified homogeneous space-time, but it
does not show that fields and background can be part of a common covariant
theory. Therefore, it is not clear whether it can rightfully be considered a
description of cosmological evolution in quantum gravity, or of quantum
space-time.

The treatment of transformations in the dressed-metric approach suffers from
several old and new problems:
\begin{itemize}
\item While deparameterization eliminates the appearance of coordinate time,
  as applied in \cite{AAN} it selects a specific reference scalar field as
  internal time (which has to be free and massless in order to play the role
  of a global measure of time). In models in which more than one choice of
  global internal time are available, quantum corrections in general imply
  inequivalent observables depending on which internal time is used
  \cite{ReducedKasner,MultChoice,TwoTimes}. Even if one does not refer to
  coordinate time, therefore, time reparameterization invariance is not
  guaranteed after quantization. This problem, which is being investigated
  with several methods, see for instance
  \cite{EffTime,EffTimeLong,EffTimeCosmo,ReducedKasner,MultChoice,QuantumRef1,QuantumRef2,QuantumRef3,QuantumRef4},
  is not specific to the dressed-metric approach and will therefore be
  disregarded here.
\item Bardeen potentials, in spite of one of their common names, are not gauge
  invariant \cite{Stewart}.\footnote{The relevance of this fact for the
    dressed-metric approach has been pointed out privately by Robert
    Brandenberger and Neil Turok.} They are invariant with respect to {\em
    small inhomogeneous} coordinate transformations in a perturbative setting,
  but they are no longer invariant if one or both of the two implied
  conditions, smallness and inhomogeneity, is violated. Curvature
  perturbations, which are available in the presence of a scalar matter field,
  are invariant provided only that coordinate transformations are small and
  not necessarily inhomogeneous, but even this condition is not met by all
  transformations relevant for perturbative cosmology: While a first-order
  description of inhomogeneity need not consider higher than first-order
  transformations, it should include large {\em homogeneous} coordinate
  changes such as a transformation from proper time to conformal time. In the
  dressed-metric approach, homogeneous coordinate transformations are
  implemented by deparameterization for the background, separately from the
  inhomogeneous sector even though they act non-trivially on Bardeen
  potentials and curvature perturbations when they are large.
\item A detailed analysis of space-time transformation in a 4-dimensional or a
  canonical setting, presented in the next section, shows that background
  transformations and those acting on perturbations do not form a direct but
  rather a semidirect product. This important algebraic structure is violated
  by the separation of background and perturbation degrees of freedom imposed
  by the dressed-metric approach, which would be compatible only with a direct
  product. As a consequence, by its very construction the dressed-metric
  approach is unable to provide the correct off-shell structure required for a
  covariant theory of background and perturbations.
\end{itemize}

\section{Space-time structure}

The perturbative form of covariance is somewhat different depending on whether
one uses a formulation of tensor fields in four dimensions or a canonical
description. However, both viewpoints lead to the same conclusion that
background and perturbative transformations form a semidirect product.

\subsection{4-dimensional formulation}

Background coordinate transformations affect only time $t$ and are generated
by vector fields of the form $f(t)\partial/\partial t$ with an arbitrary
function $f(t)$. Perturbative coordinate changes are generated by vector
fields $\xi^{\alpha}\partial/\partial x^{\alpha}$ with four components
$\xi^{\alpha}$ which are small in the sense that any products of multiple
$\xi^{\alpha}$ or of $\xi^{\alpha}$ with perturbative fields are
ignored. Bardeen potentials and curvature perturbations are constructed by
ensuring $\xi^{\alpha}$-independence of suitable combinations of metric
components, but they do not consider $f(t)$ (unless this function is small and
may be considered a contribution to $\xi^0$).

\subsubsection{Bardeen potentials and curvature perturbations}

Specifically, we may transform metric components by inserting small coordinate
changes $x^{\alpha}\mapsto x^{\alpha}+\xi^{\alpha}$ in the line element
\begin{equation} \label{LineElPert}
 {\rm d} s^2= a^2\Bigl(-(1+2\phi){\rm d}\eta^2+ 2\partial_iB{\rm
   d}\eta{\rm d} x^i 
 + \left((1-2\psi)\delta_{ij}
 + 2(\partial_i\partial_j-{\textstyle\frac{1}{3}} 
 \delta_{ij}\Delta)E\right){\rm d} x^i{\rm d} x^j\Bigr) 
\end{equation}
for linear scalar perturbations on a flat isotropic background, here using
conformal time $\eta$ and including only scalar modes. We distinguish between
time transformations, $\eta\mapsto \eta+\xi^0$, and scalar spatial
transformations, $x^i\mapsto x^i+\partial^i \xi$ with a scalar function $\xi$.
In the first case, denoting a derivative with respect to $\eta$ by a prime,
\begin{equation} \label{deta}
{\rm d}\eta^2\mapsto {\rm d}\eta^2+2\xi^{0\prime}{\rm d}\eta^2+
2\partial_i\xi^0 {\rm    d}\eta{\rm d} x^i 
\end{equation}
to first order in $\xi^0$, while $a(\eta)^2\mapsto a(\eta)^2(1+2a'\xi^0/a)$.
Rearranging the resulting line element to bring it back to the old form
(\ref{LineElPert}) but with adjusted scalar perturbations, we obtain the
transformations
\begin{equation} \label{CovTrans}
\phi\mapsto \phi+\xi^{0\prime}+\frac{a'}{a}\xi^0\quad,\quad
\psi\mapsto \psi-\frac{a'}{a}\xi^0\quad,\quad B\mapsto B-\xi^0  \quad,\quad
E\mapsto E\,.
\end{equation}
Notice that the transformation of $B$ follows only if $\partial_i\xi^0\not=0$
in (\ref{deta}) because the line element depends on $\partial_iB$ but not
directly on $B$. Therefore, for spatially constant $\xi^0$, or a small
background transformation, there is no need for $B$ to change, in contrast to
(\ref{CovTrans}). In fact, the transformation of $B$ is undetermined in this
case because $B\mapsto B-\alpha\xi^0$ would be consistent for any real
$\alpha$. This ambiguity is not relevant in the line element, which only
depends on $\partial_iB$, but it implies an ambiguity in the Bardeen
potentials, which depend directly on $B$ and not just its spatial derivatives.
Thus we obtain a distinction between background and perturbation
transformations even if both are small.

For small spatial transformations, we insert
\begin{equation} \label{spatialxi}
 \delta_{ij}{\rm d} x^i{\rm d} x^j \mapsto \delta_{ij}{\rm d}
 x^i{\rm d} x^j+ 2\partial_i\xi'{\rm 
   d}\eta{\rm d} x^i+ 2  \partial_i\partial_j\xi {\rm d} x^i{\rm d} x^j
\end{equation}
in the line element and read off
\begin{equation}
 \phi\mapsto\phi\quad,\quad \psi\mapsto\psi\quad,\quad B\mapsto
 B+\xi'\quad,\quad E\mapsto E+\xi\,.
\end{equation}
Therefore, $\phi$, $\psi$ and $B-E'$ are invariant with respect to spatial
transformations. (Again, the transformation of $B$ would be undetermined if
$\partial_i\xi=0$, but for spatial transformations we need
$\partial_i\xi\not=0$ in order to have a non-trivial
$\xi_i=\partial_i\xi\not=0$.) Since $B-E'$ changes to $B-E'-\xi^0$ by a time
transformation, the combinations
\begin{equation} \label{PhiPsi}
 \Phi:=\phi+\frac{a'}{a}(B-E')+(B-E')'
\quad\mbox{ and }\quad
\Psi:=\psi-\frac{a'}{a}(B-E')
\end{equation}
are invariant, provided $\xi^0$ is not spatially constant. 

If there is a matter scalar field, $\varphi=\bar{\varphi}+\delta\varphi$, its
perturbation transforms by
$\delta\varphi\mapsto\delta\varphi+\bar{\varphi}'\xi^0$. Therefore, one can
obtain $\xi^0$-independent combinations, the curvature perturbations
\begin{eqnarray}
 {\cal R}_1&=&\psi+\frac{a'}{a\bar{\varphi}'}\delta\varphi \label{R1}\\
 {\cal R}_2 &=& \phi-\frac{1}{2}\left(\frac{a}{a'}\right)'\psi
-\frac{1}{\bar{\varphi}'}\left(\frac{a'}{a}-
  \frac{\bar{\varphi}''}{\bar{\varphi}'}\right) \delta\varphi
+\frac{1}{2}\frac{a}{a'}\psi'- \frac{1}{2\bar{\varphi}'} \nonumber
\delta\varphi'
\end{eqnarray}
without using $B$. These perturbations, unlike Bardeen potentials, are
invariant also with respect to spatially constant $\xi^0$, but not with
respect to large background transformations.

Formulating the dressed-metric approach using curvature perturbations instead
of Bardeen potentials implies that we do not have to distinguish between small
background transformations and perturbative transformations. However, there
remain non-trivial large background transformations, hence the additional step
of deparameterization in the approach. Large background transformations change
curvature perturbations merely by reparameterizations, such as replacing
$a'/(a\bar{\varphi}')$ with $\dot{a}/\dot{\bar{\varphi}}$ when transforming
from conformal time to proper time. Formally, the approach therefore does take
into account all relevant transformations. However, the way it does so
violates the required off-shell structure of background and perturbative
transformations.

\subsubsection{Algebraic structure}
\label{s:AlgCov}

Background and perturbative transformations are not independent but
algebraically related. The commutator of two such transformations or of their
generating vector fields, given by
\begin{equation} \label{Comm}
 \left[f(t)\frac{\partial}{\partial t}, \xi^{\alpha}\frac{\partial}{\partial
     x^{\alpha}}\right]= f\dot{\xi}^{\alpha} \frac{\partial}{\partial
   x^{\alpha}}- \dot{f}\xi^0 \frac{\partial}{\partial t}\,,
\end{equation}
is a perturbative transformation. Using pairs
\begin{equation}
 \left(f, \xi^{\alpha}\right) \in {\cal V}_{\rm background} \oplus {\cal
   V}_{\rm pert} = {\cal V}
\end{equation}
of background and perturbation vector fields, arranged by perturbative order
to make the algebraic structure more clear, the combination of both types of
transformations is therefore a semidirect product:
\begin{equation}
 [(f_1,\xi_1^{\alpha}),(f_2,\xi_2^{\alpha})]= (f_1\dot{f}_2-f_2\dot{f}_1,
 \zeta^{\alpha}) 
\end{equation}
with
\begin{equation}
 \zeta^{\alpha}= f_1\dot{\xi}_2^{\alpha}- f_2\dot{\xi}_1^{\alpha}-
 \delta_{0}^{\alpha}\left(\dot{f}_1\xi_2^0-\dot{f}_2\xi_1^0\right)
\end{equation}
depending on $\xi_1$ and $\xi_2$ as well as $f_1$ and $f_2$. 

This bracket shows that ${\cal V}_{\rm background}$ is non-Abelian, with
bracket $[f_1,f_2]_{\rm background}=f_1\dot{f}_2-f_2\dot{f}_1$, while ${\cal
  V}_{\rm pert}$ is Abelian, $[\xi_1^{\alpha},\xi_2^{\alpha}]_{\rm pert}=0$.
However, the full bracket in ${\cal V}$ has an extra term $\zeta^{\alpha}$,
which can be written as $\zeta^{\alpha}= \phi(f_1)\xi_2^{\alpha}-
\phi(f_2)\xi_1^{\alpha}$ with the homomorphism
\begin{equation}
 \phi(f)\xi^{\alpha}= f\dot{\xi}^{\alpha}-\delta_0^{\alpha}\dot{f}\xi^0
\end{equation}
from ${\cal V}_{\rm background}$ to the derivations on ${\cal V}_{\rm
  pert}$. (It clearly maps to derivations because ${\cal V}_{\rm pert}$ is
Abelian. The homomorphism property can be shown by a direct calculation.)
Therefore, the bracket on ${\cal V}$ can be written as
\begin{equation}
 [(f_1,\xi_1^{\alpha}),(f_2,\xi_2^{\alpha})]= ([f_1,f_2]_{\rm background},
 [\xi_1^{\alpha}, \xi_2^{\alpha}]_{\rm pert}+\phi(f_1)\xi_2^{\alpha}-
 \phi(f_2)\xi_1^{\alpha})\,,
\end{equation}
identifying 
\begin{equation}
 {\cal V}={\cal V}_{\rm background}\ltimes_{\phi} {\cal V}_{\rm pert}
\end{equation}
as the semidirect product of the Lie algebras ${\cal V}_{\rm background}$ and
${\cal V}_{\rm pert}$.

We do not have a direct product that could be implemented by separate
treatments of invariance, such as deparameterization for the background and
curvature perturbations for the inhomogeneous fields. While the dressed-metric
approach is formally consistent in that it eliminates the relevant
transformations, it does so incorrectly by ignoring their interrelated
off-shell nature. In the next section we will demonstrate explicitly that
there is no off-shell completion of the attempted invariance proposed by the
dressed-metric approach, but first we review the off-shell structure in a
canonical setting.

\subsection{Canonical formulation}

One might think that the canonical formulation should not have a non-zero
commutator of background and perturbative transformations because fields on a
fixed spatial slice do not have any time dependence, such that the time
derivatives on the right-hand side of (\ref{Comm}) vanish. (Time dependence in
canonical transformations is not explicit but implemented by an additional
term added to the usual constraints which depends on momenta of lapse and
shift and has coefficients given by initial values of time derivatives of the
fields \cite{LapseGauge,CUP}.) However, the canonical description must be
equivalent to the 4-dimensional formulation, and therefore should give rise to
a related semidirect product of background and perturbative
transformations. The main mathematical difference is that canonical
transformations form a Lie algebroid \cite{ConsAlgebroid,ConsAlgebroid2}
rather than a Lie algebra.

\subsubsection{Algebroid}

\begin{figure}
\begin{center}
\includegraphics[width=16cm]{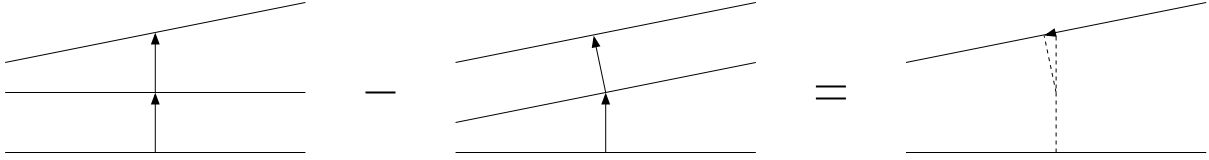}
\caption{Non-zero commutator of a homogeneous background transformation and a
  perturbative transformation (here, linear spatial dependence), equal to a
  non-zero spatial displacement. \label{Fig:SemiDir}}
\end{center}
\end{figure}

Geometrically, the product remains semidirect, not because of time derivatives
but because the canonical generators, given by constraints, refer to
directions normal to spatial slices rather than time directions determined by
a coordinate \cite{ADM}. As a consequence, a perturbative inhomogeneous
transformation changes the normal directions, such that a subsequent
background transformation acquires new directions compared with one applied
before the inhomogeneous transformation; see Fig.~\ref{Fig:SemiDir}.

The specific commutator follows from a restriction of the full
hypersurface-deformation brackets of the Hamiltonian and diffeomorphism
constraints, $H[N]$ and $D[M^a]$. We obtain background transformations by
applying the Hamiltonian constraint to homogeneous lapse functions $\bar{N}$,
while perturbative inhomogeneous constraints are obtained by specializing the
Hamiltonian constraint to a small inhomogeneous perturbation, $\delta N$, and
the diffeomorphism constraint to a small inhomogeneous vector field, $\delta
M^a$. The leading perturbative expressions are then obtained by expanding the
constraints $H[\delta N]$ and $D[\delta M^a]$ up to quadratic dependence on
the fields, counting $\delta N$ and $\delta M^a$ as first-order
contributions. The general bracket \cite{DiracHamGR}
\begin{equation} \label{HH}
 [H[N_1],H[N_2]]= D[q^{ab}(N_1\partial_bN_2-N_2\partial_bN_1)]
\end{equation}
with the inverse spatial metric $q^{ab}=a^{-2}\delta^{ab}$ then turns into
\begin{equation} \label{HHpert}
 [H[\bar{N}],H[\delta N]]= D[a^{-2} \bar{N}\partial^a\delta N]
\end{equation}
with a non-zero right-hand side. More generally,
\begin{equation} \label{HHpertbd}
 [H[\bar{N}_1+\delta N_1],H[\bar{N}_2+\delta N_2]]= 
D[a^{-2} (\bar{N}_1\partial^a\delta N_2-\bar{N}_2\partial^a\delta N_1)]
\end{equation}
while all brackets involving $D[\delta M^a]$ are zero to first perturbative
order.

In the canonical formulation, the background bracket is Abelian because
$[H[\bar{N}_1],H[\bar{N}_2]]=0$ for any spatially constant $\bar{N}_1$ and
$\bar{N}_2$. The generators of perturbative inhomogeneous transformations,
$H[\delta N]$ and $D[\delta M^a]$, also form an Abelian Lie algebra, because
the right-hand side of (\ref{HH}) vanishes to the order considered here when
both $N_1=\delta N_1$ and $N_2=\delta N_2$ are of first order. The
Lie-algebroid structure of the full bracket (\ref{HH}) therefore seems to
remain only in the non-trivial relation (\ref{HHpert}) between background and
perturbation generators. It is nevertheless possible to interpret both
background and perturbations as Lie algebroids, ${\cal E}_{\rm background}$
and ${\cal E}_{\rm pert}$, respectively, over the same base manifold $X_{\rm
  pert}$ of perturbed metrics. (Background metrics might seem sufficient for
${\cal E}_{\rm background}$, but using the same base manifold for ${\cal
  E}_{\rm background}$ and ${\cal E}_{\rm pert}$ is convenient for the
construction of a semidirect product.)

With a base manifold of metrics, (\ref{HHpertbd}) determines the algebroid
bracket only for constant sections, that is $\bar{N}$ and $(\delta N,\delta
M^a)$ which do not depend on the metric (while the perturbations $\delta N$
and $\delta M^a$ may always depend on the spatial position). If we allow
metric-dependent functions, the Lie algebroid ${\cal E}_{\rm background}$ is
no longer Abelian because the background part of the bracket (\ref{HHpertbd})
should then be generalized to
\begin{equation} \label{HHq}
 [H[\bar{N}_1],H[\bar{N}_2]] = H[\bar{N}_1 \delta_n\bar{N}_2-
 \bar{N}_2\delta_n\bar{N}_1]
\end{equation}
where $\delta_nN= (\delta N/\delta q_{ab}) {\cal L}_n q_{ab}$ is the normal
derivative of $N$, constructed by the chain rule using the Lie derivative
${\cal L}_n$ along the vector field normal to hypersurfaces. (This extension
can be derived from the Poisson bracket of Hamiltonian constraints with
metric-dependent lapse functions.)  The Lie algebroid ${\cal E}_{\rm pert}$
remains Abelian because the right-hand side of an equation analogous to
(\ref{HHq}) with $\bar{N}$ replaced by $\delta N$ would be of second order.

The anchor map of a Lie algebroid ${\cal E}$, defined as
\begin{equation}
 \rho\colon\Gamma(T{\cal E})\to \Gamma(TX)
\end{equation}
such that 
\begin{equation}
 [e_1,fe_2]= f[e_1,e_2]+(\rho(e_1)f)e_2
\end{equation}
for any $e_1,e_2\in \Gamma(T{\cal E})$ and $f\in C^1(X)$, is necessarily zero
for Abelian brackets, that is for ${\cal E}_{\rm pert}$ in our case. The
non-Abelian bracket of ${\cal E}_{\rm background}$ is compatible with the
anchor map $\bar{N}\mapsto \delta q_{ab}= \bar{N}{\cal L}_nq_{ab}$. These two
anchor maps are equivalent to the first-order perturbative content of the full
anchor, given by $(N,M^a)\mapsto \delta q_{ab}= {\cal L}_{Nn+M}q_{ab}$
\cite{ConsAlgebroid}.

Abstractly, we denote elements in the fiber of the first
Lie algebroid, ${\cal E}_{\rm background}$, simply by
$\bar{N}\in{\mathbb R}$. Elements of fiber of the second Lie algebroid,
${\cal E}_{\rm pert}$ which is Abelian, are given by $(\delta N,\delta
M^a)$, where $\delta N$ and $\delta M^a$ depend on the spatial position
and therefore form infinite-dimensional fibers. The map
\begin{equation}
 \psi(\bar{N})(\delta N,\delta M^a)= (0,a^{-2} \bar{N}\partial^a\delta N)
\end{equation}
defines a Lie algebroid morphism from ${\cal E}_{\rm background}$ to the
derivations on ${\cal E}_{\rm pert}$. (This map is well-defined because
background metrics, parameterized by the scale factor $a$, are included in
both base manifolds. It maps to derivations because ${\cal E}_{\rm pert}$ is
Abelian. In order to show the morphism property, note that $\delta_n\bar{N}$
is of first order, such that $\delta_n\bar{N}\partial^a\delta N\sim 0$ is of
second order and therefore treated as zero.)

The bracket (\ref{HHpertbd}) together with the vanishing
brackets involving spatial deformations now appear in the form
\begin{equation}
 [(\bar{N}_1,(\delta N_1,\delta M_1^a)),(\bar{N}_2,(\delta N_2,\delta
 M_2^a))]= ([\bar{N}_1,\bar{N}_2], \psi(\bar{N}_1)(\delta N_2,\delta M_2^a)-
 \psi(\bar{N}_2)(\delta N_1,\delta M_1^a))\,,
\end{equation}
which is the same as the bracket of a semidirect products of Lie algebroids
defined in \cite{LieAlgebroidExt}, where the analog of ${\cal E}_{\rm pert}$
(but not of ${\cal E}_{\rm background}$) is required to be Abelian in to avoid
obstructions. The general construction determines a semidirect product with
anchor map inherited directly from ${\cal E}_{\rm background}$, just as we
have found here. We therefore have shown that
\begin{equation}
 {\cal E}= {\cal E}_{\rm background} \ltimes_{\psi} {\cal E}_{\rm pert} \,.
\end{equation}

Comparing with the 4-dimensional perspective, although the precise algebraic
structure of canonical transformations is rather different from that found
in Section~\ref{s:AlgCov}, the bracket of a semidirect product of background
and perturbation transformations is obtained in both cases.

\subsubsection{Poisson structure}

A formal derivation of the crucial equation (\ref{HHpert}) through Poisson
brackets of phase-space representations of the hypersurface-deformation
generators shows the interplay of different perturbative orders in this
result. Following the formalism developed in
\cite{ConstraintAlgebra,ScalarGaugeInv}, or \cite{HamGaugePert} for canonical
perturbation theory in metric variables, we coordinatize the gravitational
phase space in triad form, given by the components $E^a_i$ of a densitized
triad and the corresponding components of extrinsic curvature, $K_a^i$. With
perturbative inhomogeneity, we write $E^a_i=p \delta^a_i+\delta E^a_i$ and
$K_a^i=k\delta_a^i+\delta K_a^i$, where the background variables $p$ and $k$
depend only on time and their internal frame has been fixed by choosing the
background fields to be proportional to the Kronecker delta. For simplicity,
we will assume that $p>0$, fixing the orientation of space. The background
variables can then be derived from the fields by integrating over a fixed
spatial region ${\cal V}$ of coordinate volume $V_0=\int_{\cal V}{\rm d}^3x$:
\begin{equation}
 p=\frac{1}{V_0} \int_{\cal V} E^a_i\delta_a^i{\rm d}^3x\quad,\quad
 k=\frac{1}{V_0} \int_{\cal V}K_a^i\delta^a_i{\rm d}^3x\,.
\end{equation}
In order to avoid double-counting the background variables, we impose linear
second-class constraints
\begin{equation}
  \int_{\cal V}\delta E^a_i\delta^i_a{\rm d}^3x=0=\int_{\cal V}\delta
  K_a^i\delta^a_i{\rm 
    d}^3x
\end{equation}
on the perturbation fields. With these conditions, we obtain the basic Poisson
brackets
\begin{equation}
 \{k,p\}=\frac{8\pi G}{3V_0} \quad,\quad  \{\delta K_a^i(x),\delta E^b_j(y)\}=
 \delta_a^b \delta^i_j\left(\delta(x,y)- \frac{1}{V_0}\right)\,.
\end{equation}
(The subtraction of the constant $1/V_0$ refers to the Dirac bracket of fields
subject to linear second-class constraints, but it will not contribute to the
following calculations.)

For a spatially flat isotropic model in triad form, we have the background
constraint
\begin{equation} \label{barH}
 \bar{H}= -\frac{3V_0}{8\pi G} \sqrt{p}k^2\,,
\end{equation}
the first-order constraint
\begin{equation}
 H^{(1)}[\delta N]= \frac{1}{16\pi G} \int{\rm d}^3x \delta N
 \left(-4k\sqrt{p}\delta_j^c\delta K_c^j- \frac{k^2}{\sqrt{p}}
   \delta_c^j\delta E_j^c+ \frac{2}{\sqrt{p}} \partial_c \partial^j\delta
   E_j^c\right)\,,
\end{equation}
and the second-order constraint
\begin{eqnarray}
 H^{(2)}[\bar{N}] &=& \frac{\bar{N}}{16\pi G} \int{\rm d}^3x
 \left(\sqrt{p}\delta K_c^j \delta K_d^k \delta_k^d\delta_j^d- \sqrt{p}
   \left(\delta K_c^j \delta_j^c\right)^2 - 2\frac{k}{\sqrt{p}} \delta
   E_j^c\delta K_c^j \right.\nonumber\\
&& \left. - \frac{k^2}{2p^{3/2}} \delta E_j^c \delta E_k^d
   \delta_c^k\delta_d^j + \frac{k^2}{4p^{3/2}} \left(\delta
     E_j^c\delta^c_j\right)^2 - \frac{1}{2p^{3/2}} \delta^{jk}
   \left(\partial_c \delta E^c_j\right) \left(\partial_d\delta
     E^d_k\right)\right)\,.
\end{eqnarray}
Moreover, the first-order diffeomorphism constraint is 
\begin{equation}
 D[\delta M^c] = \frac{1}{8\pi G} \int_{\cal V}{\rm d}^3x \delta
 M^c\left(p\delta_k^d \partial_c\delta K_d^k- p\partial_j \delta K_c^j-
   k\delta_c^j \partial_d \delta E_j^d\right)\,.
\end{equation}
The background diffeomorphism constraint vanishes identically, and no
second-order expression is required for our purposes.

Let us first consider only the background constraint, $\bar{N}\bar{H}$, and
the first-order constraint, $H^{(1)}[\delta N]$, in the Poisson bracket
\begin{equation}\label{HH1}
 \{\bar{N}\bar{H},H^{(1)}[\delta N]\}= \frac{1}{16\pi G} \int_{\cal V}{\rm
   d}^3x \bar{N} \delta N \left(2k^2 \delta_j^c \delta K_c^j- 2\frac{k^3}{p}
   \delta_c^j \delta E_j^c + 2\frac{k}{p} \partial_c\partial^j \delta
   E_j^c\right)\,.
\end{equation}
It is easy to see that this bracket, which is a first-order expression, is not
a linear combination of the available first-order constraints, $H^{(1)}[\delta
N]$ and $D[\delta M^c]$.  Therefore, if we combine only background and
first-order constraints, we not only fail to produce the correct bracket
(\ref{HHpert}) of perturbative hypersurface deformations but, worse, obtain
an anomalous gauge system in which the constraint brackets do not close.

This problem can easily be solved by realizing that the second-order
constraint $H^{(2)}[\bar{N}]$, while it can be ignored in the constraint
equations imposed on first-order dynamics, should be included in the
constraint brackets because its Poisson bracket with a first-order constraint
is of first order. The second-order constraint therefore contributes to
the first-order gauge flow relevant for a theory of first-order
perturbations. Indeed, the Poisson bracket
\begin{eqnarray}\label{H2H1}
 \{H^{(2)}[\bar{N}],H^{(1)}[\delta N]\}&=& \frac{1}{32\pi G} \int_{\cal V}{\rm
   d}^3x \bar{N}\left(\delta N \left(-8k^2\delta^c_j\delta K_c^j + 4\frac{k^3}{p}
   \delta_c^j \delta E_j^c\right.\right.\\
&&\left.\left.\qquad\qquad\qquad\qquad\qquad
 + 4\frac{k}{p} \partial_c \partial^j \delta E_j^c+
   4k^2\delta_j^c \delta K_c^j\right)\right.\nonumber\\
&&\left.
+ 4 \left(\delta
 K_c^j \partial_j\partial^c\delta N- \delta_j^c\delta K_c^j
 \delta_k^d\partial_d\partial^k\delta N- \frac{k}{p} \delta
 E^c_j\partial_c\partial^j\delta N\right)\right)\nonumber
\end{eqnarray}
provides just the right terms for (\ref{HH1}) and (\ref{H2H1}) to combine to
\begin{eqnarray}
 \{\bar{N}\bar{H}+H^{(2)}[\bar{N}],H^{(1)}[\delta N]\}&=& \frac{1}{8\pi G}
 \int_{\cal V}{\rm d}^3x \frac{\bar{N}\partial^c\delta N}{p}
 \left(p\delta_k^d\partial_c \delta K^j_d- p\partial_j\delta K^j_c-
   k\delta_c^j\partial_d \delta E_j^d\right)\nonumber\\
&=& D[p^{-1}\bar{N}\partial^c\delta N]\,, \label{HHpert2}
\end{eqnarray}
equivalent to (\ref{HHpert}).

With hindsight, the result of this rather technical calculation is not
surprising if one only considers that a second-order constraint can generate a
first-order gauge flow. Together with the general condition that all flows of
the same order should be included on the same footing, it is clear that one
cannot obtain an anomaly-free constrained system to first order if only the
background and first-order constraints are included. In our following
discussion it will be useful to see the presented details of how this
calculation works in order to rule out the specific proposal made in the
dressed-metric approach.

\section{The metric's new clothes}

In Riemannian geometry, the metric $g_{\alpha\beta}$ is subject to the tensor
transformation law such that the line element
\begin{equation}
 {\rm d}s^2=g_{\alpha\beta}{\rm d}x^{\alpha}{\rm d}x^{\beta}
\end{equation}
is invariant with respect to coordinate changes, ${\rm d}x^{\alpha'}=(\partial
x^{\alpha'}/\partial x^{\alpha}) {\rm d}x^{\alpha}$. The line element
therefore provides a coordinate-independent meaning of distances on which
Riemannian geometry is based. In a geometrical field theory such as general
relativity, this important condition on the metric is an off-shell property
which cannot be tested if one restricts one's attention only to solutions of
the canonical constraints or to Dirac or other observables.

If the theory is quantized canonically, coordinate transformations are
unmodified because the space-time coordinates $x^{\alpha}$ are not phase-space
functions. (We ignore here the possibility that one might wish to modify the
geometry in addition to canonically quantizing gravity, for instance by making
it non-commutative. Such a procedure would go beyond standard canonical
quantization, and it is certainly not envisioned in \cite{AAN}.) Some of the
components of $g_{\alpha\beta}$, however, represent phase-space degrees of
freedom and may therefore be subject to quantum corrections not only in their
dynamics but also in their behavior under gauge transformations. The
covariance question in canonical quantizations of gravity therefore asks
whether a quantum modified (or dressed) $\tilde{g}_{\alpha\beta}$ has {\em
  off-shell} transformations consistent with coordinate transformations. If
this question is not answered in the affirmative, the standard interpretation
of the metric through a line element is no longer available, demoting
$\tilde{g}_{\alpha\beta}$ to a purely formal object without geometrical
significance.

In \cite{AAN}, different versions of line elements have uncritically been
introduced for modified metrics without asking the covariance question. In
fact, since the formalism defined in \cite{AAN} is purely on-shell, using
deparameterization of the background dynamics together with Bardeen potentials
or curvature perturbations, it is not amenable to a direct test of
covariance. This lack of control on an important physical requirement may in
itself present a good reason to discard the dressed metric.

It is possible to go even further and show that the modified dynamics
used by the dressed-metric approach in order to obtain bouncing background
solutions cannot represent on-shell solutions of a covariant off-shell
theory. To do so, we use the canonical version of the tensor-transformation
law dual to standard coordinate transformations, given by gauge generators
subject to hypersurface-deformation brackets. As we have already seen,
perturbative inhomogeneity to first order requires us to use the Hamiltonian
constraint up to second order because a second-order contribution may well
generate a first-order flow. The dressed metric approach is half-way aware of
this important fact because it derives a dynamical flow using second-order
generators, determining the dynamical vector field
\begin{equation}
 X_{\rm Dyn}^{\alpha}= \Omega_{o}^{\alpha\beta}\partial_{\beta}S_o[N_{\rm
   hom}]+ \Omega_1^{\alpha\beta} \partial_{\beta}S_2'[N_{\rm hom}]
\end{equation}
in the notation of \cite{AAN}. The generator $S_2'$ corresponds to our
$H^{(2)}$, but is written in terms of curvature perturbations, $T_{\vec{k}}$
and their momenta $P_{\vec{k}}$, for tensor modes:
\begin{equation}
 S_2'[a^3\ell^3/p_{\phi}]= \frac{1}{2}\sum_{\vec{k}}
 \left(4\frac{\kappa}{p_{\phi}} |P_{\vec{k}}|^2+ \frac{k^2}{4\kappa}
   \frac{a^4}{p_{\phi}} |T_{\vec{k}}|^2\right)
\end{equation}
using the choice of lapse function, $N_{\rm hom}= a^3\ell^3/p_{\phi}$,
preferred in \cite{AAN}.  Here, $p_{\phi}$ is the constant background momentum
of the free, massless scalar field used for deparameterization, while
$\kappa=8\pi G$ and $\ell$ is a length parameter that is not relevant for our
purposes.

Quantization is then performed separately for $S_o$ and $S_2'$. The background
generator $S_o$, or our $\bar{H}$, is modified by loop quantization, replacing
its quadratic momentum dependence in (\ref{barH}) with a bounded
function. (The precise form of this modification does not matter for the
arguments given below.) The perturbation part $S_2'$, however, remains
quadratic in momenta and has only slightly modified coefficients,
\begin{equation} \label{S2p}
  \tilde{S}_2'[a^3\ell^3/p_{\phi}]= \frac{1}{2}\sum_{\vec{k}}
  \left(4\kappa\langle\hat{p}_{\phi}^{-1}\rangle |P_{\vec{k}}|^2+
    \frac{k^2}{4\kappa} 
    \langle \hat{p}_{\phi}^{-1/2}\hat{a}^4\hat{p}_{\phi}^{-1/2}\rangle
    |T_{\vec{k}}|^2\right) \,,
\end{equation}
where background operators are reduced to (internal) time-dependent functions
by taking expectation values in a background state. The same expectation
values are then used to define a dressed metric in the proposed line element
\begin{equation} \label{Dressed}
{\rm d}\tilde{s}^2= \tilde{g}_{ab} {\rm d}x^a{\rm d}x^b=
- \ell^6 \langle\hat{p}_{\phi}^{-1}\rangle^{1/2}\langle
\hat{p}_{\phi}^{-1/2}\hat{a}^4\hat{p}_{\phi}^{-1/2}\rangle^{3/2} {\rm
  d}\phi^2+ \langle\hat{p}_{\phi}^{-1}\rangle^{-1/2} \langle
\hat{p}_{\phi}^{-1/2}\hat{a}^4\hat{p}_{\phi}^{-1/2}\rangle^{1/2} {\rm
  d}\vec{x}^2\,,
\end{equation}
such that the coefficients in (\ref{S2p}) correspond to the classical
expression if one were to use the dressed metric to compute it. (The proposal
in \cite{AAN} also includes a metric operator such that
\begin{equation}
{\rm d}\hat{s}^2= \hat{g}_{ab} {\rm d}x^a{\rm d}x^b=
- \ell^6 \hat{p}_{\phi}^{-1}\hat{a}^6\hat{p}_{\phi}^{-1} {\rm
  d}\phi^2+ \hat{a}^2 {\rm  d}\vec{x}^2\,.
\end{equation}
However, since geometrical procedures do not measure operators, this object
does not have any well-defined meaning, other than that it produces
(\ref{Dressed}) as a formal expectation value.)

The coefficients of the dressed metric are background functions and are
therefore modified if one inserts solutions of the holonomy-modified
background constraint. Moreover, there are state-dependent quantum corrections
in these coefficients, defined through expectation values, which could be
derived systematically in a moment expansion in the framework of effective
canonical constraints; see for instance \cite{EffAc,EffCons,EffConsRel}.
However, these two quantum corrections cannot counter modifications of the
background constraint so as to make the bracket (\ref{HHpert}) work out, for
the following reasons:
\begin{itemize}
\item The off-shell behavior of the metric does not depend on what kind of
  background solutions are entered, and therefore it does not know about
  holonomy modifications.  For the off-shell behavior, relevant for
  covariance, coefficients in (\ref{Dressed}) depending on $a$ and $p_{\phi}$
  (and possibly their moments) are merely phase-space coordinates, just like
  the corresponding functions in the modified background constraint. The
  off-shell theory of the dressed-metric approach therefore corresponds to a
  system in which only the background constraint, $\bar{H}$, has been modified
  by using holonomies, but not the second-order constraint,
  $H^{(2)}$. Moreover, also the first-order constraint, $H^{(1)}$ is
  unmodified because \cite{AAN} uses the classical curvature perturbations
  without modifications that would result if gauge transformations generated
  by $H^{(1)}$ were modified; see \cite{ScalarGaugeInv,ScalarHolInv}. The
  bracket (\ref{H2H1}) then remains unchanged while (\ref{HH1}) is modified,
  eliminating important cancellations that led to the combined result
  (\ref{HHpert2}). The dressed-metric approach functions by modifying only the
  background constraint, making it impossible to realize a valid version of
  the perturbative hypersurface-deformation bracket (\ref{HHpert}).
\item If moments of a state that result from a systematic semiclassical
  expansion of the expectation values in (\ref{S2p}) were to counter the
  background modification, they would have to be fixed, severely restricting
  the class of quantum states that are allowed to propagate. Even if there
  were moments such that the bracket (\ref{HHpert}) could be closed after
  background modifications, the resulting mismatch of classical and quantum
  degrees of freedom would amount to an anomaly. (Recall that an anomaly in a
  constraint system implies that the system becomes over-constrained, imposing
  an additional constraint such as $\{\bar{N}\bar{H},H^{(1)}[\delta N]\}=0$ if
  the left-hand side is no longer zero on the solutions space of the original
  constraints.)
\end{itemize}

In addition to violating covariance, the dressed metric has the following
problem: The dressed metric depends on the ordering chosen for operators in
the expectation-value components. Moreover, for different background gauges,
corresponding to different phase-space function for the background lapse
$\bar{N}$, different operator products appear, giving rise to different
ordering ambiguities. Therefore, choosing a different background gauge in
general results in an inequivalent dressed metric.  Ordering issues can
potentially be ignored if one uses sharply peaked states, such that
fluctuation terms are negligible. Such an assumption is sometimes suggested by
the dressed-metric approach, as in ``one knowns that there exist background
quantum geometries $\Psi_o$ which are {\em very} sharply peaked'' (emphasis in
\cite{AAN}). However, this assumption is not justified in the Planck regime
\cite{Infrared,EFTLQC}, where a dressed metric would be most relevant.

\section{Conclusions}

Covariance in canonical quantum gravity is a subtle issue. It requires a
formulation of quantum effects such that the classical
hypersurface-deformation brackets (\ref{HH}) are obtained in the classical
limit of the theory, while a closed, anomaly-free set of brackets is realized
for non-zero $\hbar$ which vanishes when the constraints are solved but is not
necessarily of the classical form. This statement includes two conditions,
which cannot always both be met. For instance, a possible Abelianization of
the bracket in some midisuperspace models \cite{LoopSchwarz,LoopSchwarz3}
always leads to anomaly-free quantum constraints but even then is not
guaranteed to be compatible with covariance
\cite{SphSymmCov,GowdyCov}. Although such quantum theories in the latter case
are formally consistent as quantizations of constrained systems, they cannot
be interpreted as models of quantum space-time because there is no
well-defined sense in which they are covariant.

As an alternative to realizations of the hypersurface-deformation brackets,
analog actions in space-time tensor form, such as certain scalar-tensor
theories, have been proposed as a possible way to demonstrate
covariance. However, while such analog actions may work in simple, isotropic
models with a small number of degrees of freedom, in all known cases they fail
to describe anisotropic models or perturbative inhomogeneity correctly. For
instance, the Palatini-$f(R)$ model proposed in \cite{ActionRhoSquared},
claimed to show that loop quantum cosmology is covariant, is equivalent to a
scalar-tensor theory with a non-dynamical scalar \cite{PalatinifR} in which
any correction to general relativity amounts to a simple cosmological constant
in vacuum models. It therefore cannot possibly describe holonomy modifications
in anisotropic vaccum models, ruling it out as a possible covariant version of
loop quantum cosmology. More recent analog actions
\cite{LimCurvLQC,HigherDerivLQC} based on mimetic gravity \cite{Mimetic} again
work in isotropic models but fail to describe anisotropies or perturbative
inhomogeneity correctly \cite{MimeticLQC,MimeticLQCPert,DefSchwarzschild2}.

As shown here, the dressed-metric approach again fails to provide a covariant
version of perturbative inhomogeneity in loop quantum cosmology, in particular
in the presence of holonomy modifications of the background dynamics that may
make it possible to have bouncing solutions. Although we have focused on the
specific formulation described in \cite{AAN} for technical details of the
constructions, similar arguments apply to related (or precursor) formulations
in \cite{QFTCosmo,QFTCosmoClass} or the ``hybrid'' approach
\cite{Hybrid,Hybrid2,HybridFlat} which share with the dressed-metric approach
the crucial feature of separating the background degrees of freedom from
inhomogeneous modes, making it impossible to implement the key relation
(\ref{HHpert}) which belongs to a semidirect product of Lie algebroids.

Our result adds to mounting evidence that models of loop quantum gravity
cannot be covariant without drastic modifications of space-time structure; see
also \cite{TransComm}.  It is sometimes suggested that a non-covariant model
which implements some quantum effects in an otherwise consistent way may be
useful as a ``first approximation'' to a complicated formulation of
cosmological dynamics in full quantum gravity. However, violating an important
consistency condition such as covariance is not an approximation at all
because it usually gives rise to uncontrolled, spurious solutions that
overshadow the relevant behavior, or to over-constrained dynamics. (See also
\cite{GaugeInvTransPlanck} for a similar result in a different setting.) As an
example, covariant versions of holonomy-modified models of loop quantum
gravity, derived in \cite{JR,ScalarHolInv,LTBII,HigherSpatial,SphSymmOp},
generically imply signature change at Planckian density. The would-be bounce
is then a 4-dimensional Euclidean region in which no deterministic evolution
exists \cite{SigImpl,Loss}. Non-deterministic behavior is an example for an
effect that cannot be considered a small correction to modified but still
deterministic dynamics, even if the modes used to determine the structure of
space-time and propagation properties are perturbative.

\section*{Acknowledgements}

This work was supported in part by NSF grant PHY-1912168. The author thanks
Jakub Mielczarek for discussions.


\end{document}